\documentclass[aps,prb,reprint,superscriptaddress]{revtex4-2}
\usepackage{graphicx}
\usepackage{amsmath}
\usepackage{amssymb}
\usepackage{color}
\usepackage{hyperref}

\begin{document}
\title{Carrier-induced transition from antiferromagnetic insulator to ferromagnetic metal in the layered phosphide EuZn$_2$P$_2$}
\author{Xiyu Chen}
\affiliation{Key Laboratory of Quantum Materials and Devices of Ministry of Education, School of Physics, Southeast University, Nanjing 211189, China}
\author{Wuzhang Yang}
\affiliation{School of Science, Westlake University, Hangzhou 310024, China}
\affiliation{Institute of Natural Sciences, Westlake Institute for Advanced Study, Hangzhou 310024, China}
\author{Jia-Yi Lu}
\affiliation{School of Physics, Interdisciplinary Center for Quantum Information and State Key Laboratory of Silicon and Advanced Semiconductor Materials, Zhejiang University, Hangzhou 310058, China}
\author{Zhiyu Zhou}
\affiliation{Key Laboratory of Quantum Materials and Devices of Ministry of Education, School of Physics, Southeast University, Nanjing 211189, China}
\author{Zhi Ren}
\affiliation{School of Science, Westlake University, Hangzhou 310024, China}
\affiliation{Institute of Natural Sciences, Westlake Institute for Advanced Study, Hangzhou 310024, China}
\author{Guang-Han Cao}
\affiliation{School of Physics, Interdisciplinary Center for Quantum Information and State Key Laboratory of Silicon and Advanced Semiconductor Materials, Zhejiang University, Hangzhou 310058, China}
\affiliation{Collaborative Innovation Centre of Advanced Microstructures, Nanjing University, Nanjing 210093, China}
\author{Shuai Dong}
%\email{sdong@seu.edu.cn}
\affiliation{Key Laboratory of Quantum Materials and Devices of Ministry of Education, School of Physics, Southeast University, Nanjing 211189, China}
\author{Zhi-Cheng Wang}
\email{wzc@seu.edu.cn}
\affiliation{Key Laboratory of Quantum Materials and Devices of Ministry of Education, School of Physics, Southeast University, Nanjing 211189, China}
\date{\today}

\begin{abstract}
EuZn$_2$P$_2$ was reported to be an insulating antiferromagnet with $T_\mathrm{N}$ of 23.5 K. In this study, single crystals of EuZn$_2$P$_2$ exhibiting metallic behavior and a ferromagnetic order of 72 K ($T_\mathrm{C}$) are successfully synthesized via a salt flux method. The presence of hole carriers induced by the Eu vacancies in the lattice is found to be crucial for the drastic changes in magnetism and electrical transport. The carriers mediate the interlayer ferromagnetic interaction, and the coupling strength is directly related to $T_\mathrm{C}$, as evidenced by the linear dependence of $T_\mathrm{C}$ and the fitted Curie-Weiss temperatures on the Eu-layer distances for ferromagnetic Eu$M_2X_2$ ($M$ = Zn, Cd; $X$ = P, As). The ferromagnetic EuZn$_2$P$_2$ shows conspicuous negative magnetoresistance (MR) near $T_\mathrm{C}$, owing to strong magnetic scattering. The MR behavior is consistent with the Majumdar-Littlewood model, indicating that the MR can be enhanced by decreasing the carrier density. Our findings suggest that Eu$M_2X_2$ has highly tunable magnetism and charge transport, making it a promising material family for potential applications in spintronics.

\end{abstract}
\maketitle

Eu-based layered compounds of Eu$M_2X_2$ ($M$ = Zn, Cd; $X$ = P, As, Sb) with a trigonal CaAl$_2$Si$_2$-type structure attract great research attention for the successive discoveries of exciting phenomena. EuCd$_2$As$_2$ was claimed to be a magnetic Weyl semimetal in the polarized state~\cite{huaDiracSemimetalTypeIV2018,sohIdealWeylSemimetal2019a,WangLL_PRB_2019,rahnCouplingMagneticOrder2018,maSpinFluctuationInduced2019,xuUnconventionalTransverseTransport2021}, and dramatic alterations in the magnetic ground state and charge transport were observed by applying pressure or changing the crystal growth condition~\cite{gatiPressureinducedFerromagnetismTopological2021,duConsecutiveTopologicalPhase2022,joManipulatingMagnetismTopological2020,roychowdhuryAnomalousHallConductivity2023,taddeiSinglePairWeyl2022}. EuCd$_2$P$_2$ shows a colossal magnetoresistance (CMR) effect due to the strong magnetic fluctuations well above the N\'{e}el temperature ($T_\mathrm{N}$) at 11 K~\cite{wangColossalMagnetoresistanceMixed2021}, and the onset of ferromagnetic (FM) order was discovered in the temperature range of the strongest CMR~\cite{sunkoSpincarrierCouplingInduced2023,homesOpticalPropertiesCarrier2023,Zhang2023}. The insulator-to-metal transition as well as the topological phase transition was observed for pressurized EuZn$_2$As$_2$~\cite{luoColossalMagnetoresistanceTopological2023}. Moreover, CMR effect is also reported for EuZn$_2$P$_2$ with semiconducting behavior recently~\cite{krebberColossalMagnetoresistanceMathrm2023}.

However, the physical properties reported with samples grown by different recipes are not consistent. For example, the perspective that EuCd$_2$As$_2$ is a topological semimetal was challenged by fresh experimental and theoretic evidence, that indicates  EuCd$_2$As$_2$ is in fact a magnetic semiconductor~\cite{santos-cottinEuCdMagneticSemiconductor2023,shiAbsenceWeylNodes2023}. In addition, insulating behavior of EuZn$_2$P$_2$ was also reported in previous experiments~\cite{berryTypeAntiferromagneticOrder2022,singhSuperexchangeInteractionInsulating2023a}. The varied measurement results suggest the properties of materials in the Eu$M_2X_2$ family are extremely sensitive to carrier concentration, which is usually induced unintentionally by vacancies in the sample.

In this study, we report the successful synthesis of single crystals of EuZn$_2$P$_2$ ($T_\mathrm{C} = 72$ K), EuZn$_2$As$_2$ ($T_\mathrm{C} = 42$ K), and EuCd$_2$P$_2$ ($T_\mathrm{C} = 47$ K) with a FM ground state. Comprehensive characterizations of magnetism and electrical transport for FM-EuZn$_2$P$_2$ are presented in the main text. We conclude that the heavy hole doping in FM-EuZn$_2$P$_2$ resulting from the Eu vacancies ($\sim5\%$) is responsible for the interlayer FM coupling, which leads to a transition from an antiferromagnetic (AFM) insulator to the FM metal. We find that these FM-Eu$M_2X_2$, including FM-EuCd$_2$As$_2$ ($T_\mathrm{C} = 26$ K) reported before~\cite{joManipulatingMagnetismTopological2020}, show a linear relationship between the Curie-Weiss temperatures $\theta_\mathrm{CW}$ (as well as $T_\mathrm{C}$) and the Eu-layer distances. This observation not only indicates the prominent role of interlayer Eu-Eu interaction in $T_\mathrm{C}$, but also suggests the probability of $T_\mathrm{C}$ promotion by decreasing the layer distance. The switchable magnetic states and charge transport behaviors make EuZn$_2$P$_2$ an excellent candidate for future spintronics.

Single crystals of FM-EuZn$_2$P$_2$ were grown via a molten salt flux, similar to the growth of FM-EuCd$_2$As$_2$~\cite{joManipulatingMagnetismTopological2020}. The details related to the sample growth can be found in the Supplemental Material (SM)~\cite{suppmatt}.  Figure~\ref{F1} displays the x-ray diffraction (XRD) pattern of FM-EuZn$_2$P$_2$ single crystal. Only sharp (00$l$) diffraction peaks were observed, indicating the high quality of the single crystals. Typically, the crystals grow as millimeter-sized hexagonal thin flakes, as shown in the left inset. The right inset illustrates the structure of EuZn$_2$P$_2$ that comprises alternating layers of triangular Eu$^{2+}$ lattice connected by the layer of edge-sharing ZnP$_4$ tetrahedra. The crystallographic $c$ axis, i.e., the distance of Eu layers, is calculated to be 7.008(5) \AA\ with (00$l$) reflections. To further analyze the crystal structure of FM-EuZn$_2$P$_2$, the single-crystal XRD (SCXRD) data were collected at $T$ = 150 K. The refined result (see Table \ref{Tab-1}) confirms the CaAl$_2$Si$_2$-type structure of FM-EuZn$_2$P$_2$, and reveals 5\% Eu vacancies in the lattice, consistent with a chemical composition resulting from energy dispersive x-ray spectroscopy (EDX, Table S1 in SM)~\cite{suppmatt}. Compared to the structural parameters reported for AFM-EuZn$_2$P$_2$ at 213 K~\cite{berryTypeAntiferromagneticOrder2022}, FM-EuZn$_2$P$_2$ has a smaller unit cell due to the lower experimental temperature, but its $c/a$ ratio is slightly larger (1.7156 vs. 1.7141 for AFM-EuZn$_2$P$_2$), agreeing with the $c$-axis expansion caused by the hole doping. In a very recent study on AFM-EuZn$_2$P$_2$, the interlayer AFM coupling was explained as realized by the superexchange interaction through the Eu-P-P-Eu path~\cite{singhSuperexchangeInteractionInsulating2023a}. Since the Eu-P and P-P bond lengths as well as the Eu-P-P bond angle are basically unchanged, and no vacancies are found for Zn$_2$P$_2$ layers of FM-EuZn$_2$P$_2$, it is plausible to assume that the interlayer AFM interaction is kept. Hence, the emerging ferromagnetism should be ascribed to the interlayer FM coupling mediated by the hole carriers. Although the substitution of Na$^+$/K$^+$ for Eu$^{2+}$ will also result in hole doping, we conclude that the carriers are mainly induced by Eu vacancies, since Na/K is absent from EDX and the heterovalent substitution in Eu$M_2X_2$ seems to be hard in a previous report~\cite{kuthanazhiMagnetismPhaseDiagrams2023}. In addition, the lattice strain caused by the minor contraction of unit cell ($\sim$0.5\%) is not likely to be cause of the alterations of properties either, which is evidenced by a recent study about the pressure effect on insulating EuCd$_2$As$_2$\cite{Chen2023}.

\begin{figure}
	\includegraphics[width=0.45\textwidth]{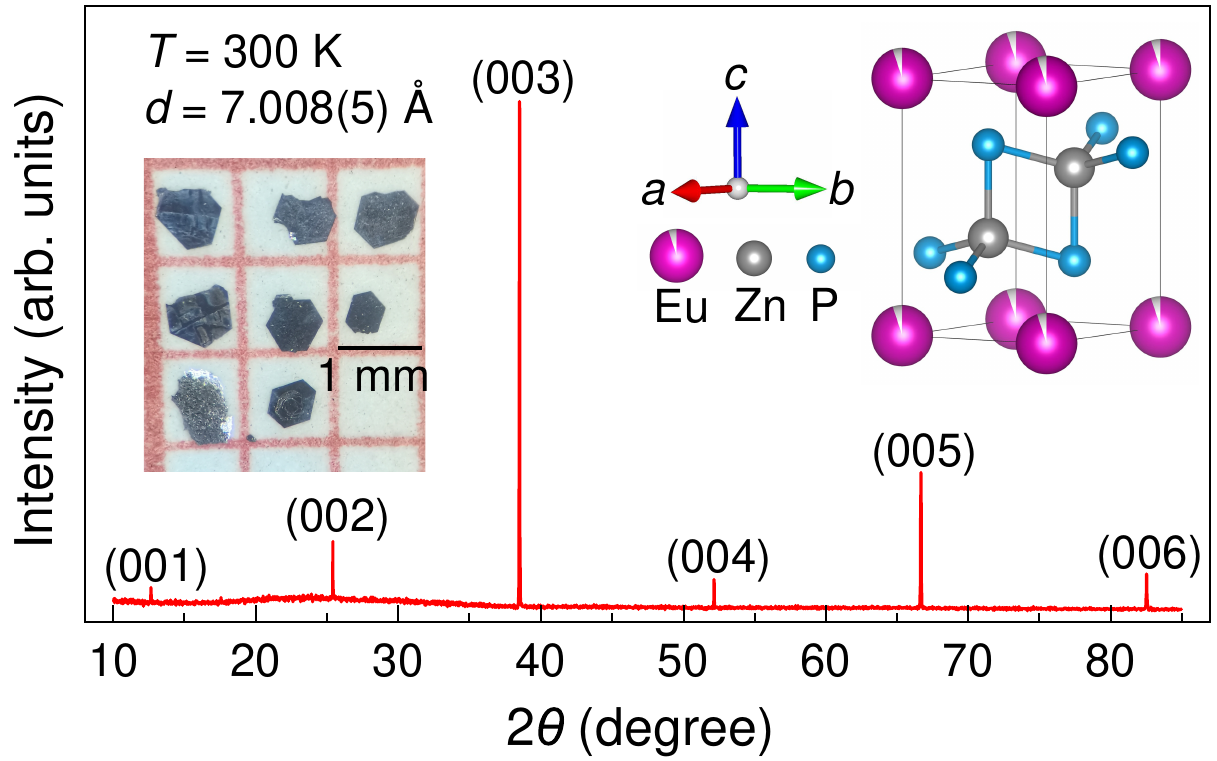}
	\caption{XRD pattern ($\theta-2\theta$ scan) of the FM-EuZn$_2$P$_2$ single crystal showing only (00$l$) reflections. Left inset shows the shape and size of typical FM-EuZn$_2$P$_2$ single crystals, and the right inset shows the crystal structure of EuZn$_2$P$_2$ with Eu vacancies.}
	\label{F1}
\end{figure}

	%%%%%%%%%%%%%%%%%%% TABLE1 %%%%%%%%%%%%%%%%%%%%%
\begin{table}
	\caption{Crystallographic data and refinement result of FM-EuZn$_2$P$_2$ from the SCXRD at 150 K~\cite{ccdcEZP}. The occupancies of Zn and P were fixed to 1.0 to avoid unphysical values greater than 1.}
	\begin{ruledtabular}\label{Tab-1}
		\begin{tabular}{ll}	
		Material & FM-EuZn$_2$P$_2$ \\
			\hline
			Crystal system  & Trigonal \\	
			Space group & $P\bar3m1$ (No. 164)  \\   
			$a$ (\AA) & 4.0765(2) \\                               
			$c$ (\AA) & 6.9936(4) \\
			$c/a$ & 1.7156 \\
%			$\alpha$ = $\beta$  & 90$\rm^o$ \\
%			$\gamma$ & 120$\rm^o$   \\    
			$V$ (\AA$^3$) & 100.648(11) \\                           
			$Z$ & 1 \\
			Eu site occupancy & 0.950(8)  \\
			Temperature (K) & 150 \\
			Radiation & Mo $K\alpha$ \\
			Reflections collected & 1424 \\
			Independent reflections & 104  \\
			$R\rm_{int}$ & 0.0819  \\
			Goodness-of-fit & 1.345 \\
			$R_1$\footnote{$R_1=\Sigma||F_o|-|F_c||/\Sigma|F_o|$.} & 0.0339 \\
			$wR_2$\footnote{$wR_2=[\Sigma w(F_o^2-F_c^2)^2/\Sigma w(F_o^2)^2]^{1/2}$.}  & 0.0984 \\	
			Fractional coordinates & \\
			Eu & (0,0,0)   \\                            
			Zn & (1/3,2/3,0.6305(4))  \\
			P & (1/3,2/3,0.2691(7))  \\			
		\end{tabular}
	\end{ruledtabular}
\end{table}

\begin{figure*}
	\includegraphics[width=0.98\textwidth]{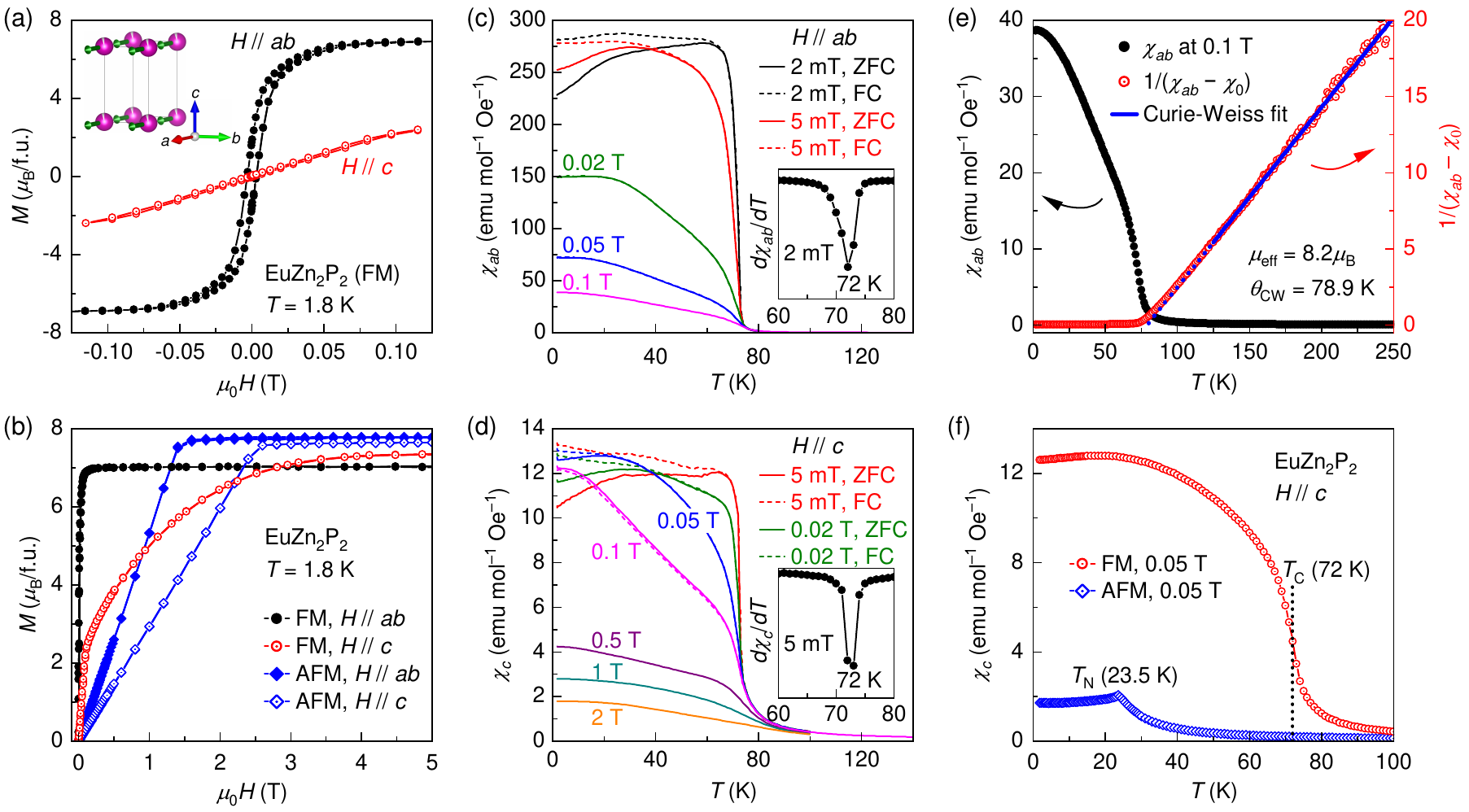}
	\caption{(a) Magnetic hysteresis loops of FM-EuZn$_2$P$_2$ at 1.8 K for fields in the $ab$ plane (black) and along the $c$ axis (red). Inset shows the spin configuration of FM-EuZn$_2$P$_2$. (b) Magnetization as a function of field for FM- (black and red circles) and AFM- (blue diamonds) EuZn$_2$P$_2$ at $T$ = 1.8 K. [(c, d)] Temperature dependences of magnetic susceptibility for field parallel to the $ab$ plane (c) and the $c$ axis (d). Their insets present the derivative of the magnetic susceptibility ($d\chi/dT$) to show the transition temperatures. (e) Curie-Weiss analysis from 125 to 300 K with $\chi_{ab}(T)$ data at 0.1 T. Only data below 250 K are shown. (f) The magnetic transitions for FM- (red) and AFM-EuZn$_2$P$_2$ (blue) at 0.05 T.}
	\label{F2}
\end{figure*}

The magnetism of FM-EuZn$_2$P$_2$ is summarized in Fig. \ref{F2}. A clear hysteresis loop can be found for the $M(H)$ curve with the in-plane field ($H\parallel ab$) at 1.8 K, shown in panel (a). And the hysteresis loop with the out-of-plane field is indistinctive ($H\parallel c$) compared to the in-plane curve, suggesting that the spins should be aligned in the $ab$ plane, as illustrated by the inset. Recent studies have demonstrated that the magnetic ordering of AFM-EuZn$_2$P$_2$ is A type, i.e. in-plane FM coupling and out-of-plane AFM coupling, whose $M(H)$ curves do not exhibit any hysteresis~\cite{berryTypeAntiferromagneticOrder2022,singhSuperexchangeInteractionInsulating2023a,krebberColossalMagnetoresistanceMathrm2023}. The significant difference in $M(H)$ curves between AFM- and FM-EuZn$_2$P$_2$ indicates that the intrinsic FM-EuZn$_2$P$_2$ is successfully prepared by changing the growth condition.

Figure \ref{F2}(b) displays the comparison of $M(H)$ curves between FM- and AFM-EuZn$_2$P$_2$. For FM-EuZn$_2$P$_2$, the $M$($H$) curves exhibit strong magnetocrystalline anisotropy. The saturation field ($H_\mathrm{sat}$) for the in-plane magnetization is only about 0.1 T, while the out-of-plane magnetization increases rapidly below 0.1 T and finally saturates at the field over 2.5 T. The similar metamagnetic transition around 0.1 T with the $c$-axis field was also reported for FM-EuCd$_2$As$_2$, which was attributed to the increasing canting of spins in the external field~\cite{roychowdhuryAnomalousHallConductivity2023}. The large magnetocrystalline anisotropy (over 25 times on the basis of $H_\mathrm{sat}$) suggests that the Eu layer (the $ab$ plane) is the magnetic easy plane. We notice that the values of $H_\mathrm{sat}$ ($H\parallel c$) for FM- and AFM-EuZn$_2$P$_2$ are comparable, while $H_\mathrm{sat}$ of FM-EuCd$_2$As$_2$ for $H\parallel ab$ is noticeably reduced. The conspicuous change in $H_\mathrm{sat}$ manifests that the FM coupling between the Eu layers is much enhanced by the induced carries. In addition, the saturated magnetizations ($M_\mathrm{sat}$) of FM-EuZn$_2$P$_2$ reach about $7 \mu_\mathrm{B}$ for both directions at 1.8 K, consistent with the Eu$^{2+}$ oxidation state.

Figures \ref{F2}(c) and 2(d) show the curves of temperature-dependent magnetic susceptibility of FM-EuZn$_2$P$_2$ under the in-plane and out-of-plane magnetic fields, respectively. Both $\chi_{ab}(T)$ and $\chi_{c}(T)$ increase dramatically below 80 K, and evident bifurcations of zero-field-cooling (ZFC) and field-cooling (FC) data are observed with small fields, which is the typical feature for ferromagnets and is consistent with the hysteresis loops in Fig. \ref{F2}(a). The splittings are rapidly suppressed with the increasing field. The Curie temperature ($T_\mathrm{C}$) of FM-EuZn$_2$P$_2$ is determined by the susceptibility derivatives ($d\chi/dT$), shown in the insets of Figs. \ref{F2}(c) and 2(d). $T_\mathrm{C}$ is identified to be 72 K by both $d\chi_{ab}/dT$ and $d\chi_{c}/dT$. Moreover, we notice that $\chi_{ab}$ is about an order of magnitude larger than $\chi_c$ for the small fields, which confirms the strong magnetocrystalline anisotropy of FM-EuZn$_2$P$_2$ in Figs. \ref{F2}(a) and 2(b).

The Curie-Weiss analysis of $\chi_{ab}(T)$ is presented in Fig. \ref{F2}(e). The data above 125 K is fitted with $\chi_{ab} = \chi_0 + C/(T - \theta_\mathrm{CW})$, which yields $C=8.47$ emu K mol$^{-1}$ Oe$^{-1}$ (effective moment $\mu_\mathrm{eff}=8.2\mu_\mathrm{B}/\mathrm{f.u.}$),  $\theta_\mathrm{CW}$ = 78.9 K, and $\chi_0 = 1.39\times10^{-4}$ emu mol$^{-1}$ Oe$^{-1}$. The value of $\mu_\mathrm{eff}$ is coincident with the theoretical value of $7.94\mu\rm_B$ for Eu$^{2+}$. The Curie-Weiss temperature $\theta_\mathrm{CW}$ is close to the experimentally determined $T_\mathrm{C}$, indicating that EuZn$_2$P$_2$ is manipulated to be an intrinsic ferromagnet rather than a canted antiferromagnet. $\chi_{c}(T)$ of AFM- and FM-EuZn$_2$P$_2$ are compared in Fig. \ref{F2}(f). AFM-EuZn$_2$P$_2$ exhibits a conspicuous peak at 23.5 K, agreeing well with the reported N\'{e}el temperature ($T_\mathrm{N}$)~\cite{berryTypeAntiferromagneticOrder2022,singhSuperexchangeInteractionInsulating2023a,krebberColossalMagnetoresistanceMathrm2023}. The distinct behaviors of $\chi_{c}(T)$ confirm the alteration of ground-state and the dominant FM interaction in FM-EuZn$_2$P$_2$. $T_\mathrm{C}$ of FM-EuZn$_2$P$_2$ is significantly higher than those of the sister materials with a FM state (47 K for EuCd$_2$P$_2$, 26 K for EuCd$_2$As$_2$, 42 K for EuZn$_2$As$_2$, see SM)~\cite{suppmatt}, which is attributed to its smaller distance of Eu layers and is discussed later.

\begin{figure}
	\includegraphics[width=0.48\textwidth]{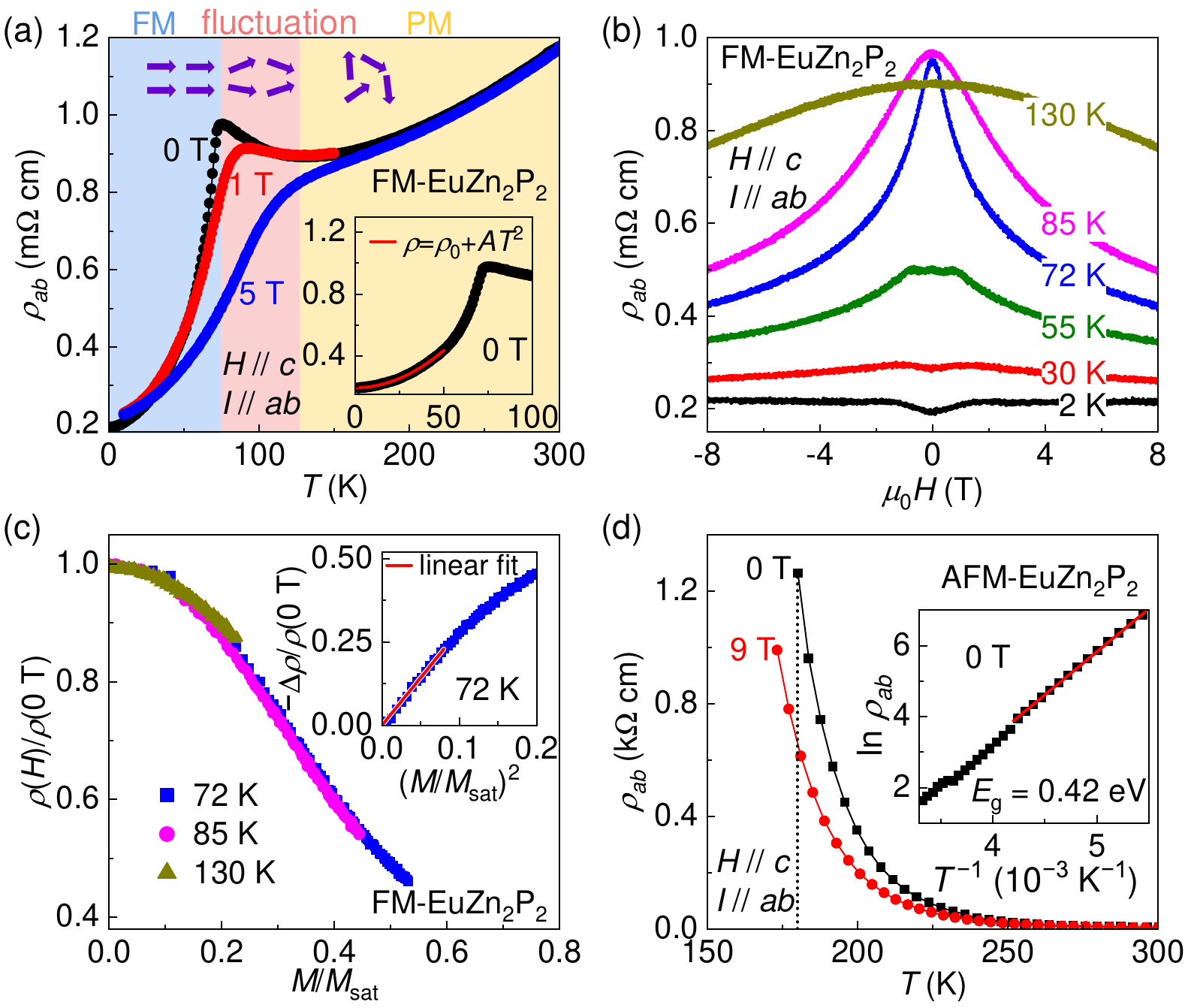}
	\caption{(a) In-plane resistivity of FM-EuZn$_2$P$_2$ as a function temperature under several fields. Inset shows a fit (red curve) with the parabolic function $\rho(T)=\rho_0+AT^2$ to the zero-field resistivity below 50 K. (b) Field-dependent resistivity of FM-EuZn$_2$P$_2$ at various temperatures from 2 to 130 K. (c) Normalized resistivity $\rho(H)/\rho(0\ \mathrm{T})$ at 72 K (blue squares), 85 K (magenta circles), and 130 K (dark yellow triangles) plotted against normalized field-induced magnetization $M/M_\mathrm{sat}$, where $M_\mathrm{sat}$ is the saturated magnetization, i.e., 7$\mu_\mathrm{B}$. Inset plots $-$MR as a function of $(M/M_\mathrm{sat})^2$ and a linear fit (red line) for $(M/M_\mathrm{sat})^2\le0.08$. (d) In-plane resistivity of AFM-EuZn$_2$P$_2$ at 0 and 9 T. Inset displays the Arrhenius plot ($\ln\rho_{ab}$ vs $T^{-1}$) of zero-field data, as well as a linear fit (red line).}
	\label{F3}
\end{figure}

FM-EuZn$_2$P$_2$ exhibits a metallic behavior, as shown in Fig. \ref{F3}(a). In the paramagnetic (PM) region (above 120 K), the zero-field in-plane resistivity ($\rho_{ab}$) decreases as the temperature goes down, and then shows a mild rise in the region of magnetic fluctuation ($T_\mathrm{C}\lesssim T\lesssim1.5T_\mathrm{C}$) due to the enhanced scattering. Finally, $\rho_{ab}$ decreases monotonically below the FM ordering at 72 K. The residual resistivity ratio (RRR, a ratio of $R_\mathrm{300 K}$ and $R_\mathrm{0 K}$) is 6, and remains almost unchanged in the field. Considering 5\% Eu$^{2+}$ vacancies in the lattice, the hole concentration in FM-EuZn$_2$P$_2$ is estimated to be $10^{21}$ cm$^{-3}$ based on the single-band model, which results in the Mott-Ioffe-Regel (MIR) limit of 0.7 m$\Omega$ cm~\cite{Gurvitch1981}. Since the residual resistivity at 2 K ($\sim$0.2 m$\Omega$ cm) is below the MIR limit, FM-EuZn$_2$P$_2$ is a good metal with a relatively long mean free path. We found that the behavior of zero-field $\rho_{ab}(T)$ below 50 K could be well described with a simple quadratic  function $\rho(T)=\rho_0+AT^2$, where the $T^2$ dependence is contributed by the electron-electron or electron-magnon scattering. The resultant fitting parameters are $\rho_0=0.193$ m$\Omega$ cm, $A=9.60\times10^{-5}$ m$\Omega$ cm K$^{-2}$. The typical Fermi liquid behavior indicates the negligible electron-phonon scattering at low temperatures and the weak electron correlation of FM-EuZn$_2$P$_2$.

The resistivity peak at $T_\mathrm{C}$ is suppressed by applying the external magnetic fields. The curves of $\rho_{ab}$ as a function of field are plotted in Fig. \ref{F3}(b). The maximum negative magnetoresistance (MR) is achieved at $T_\mathrm{C}$, which is over $-50\%$ at 8 T with the definition $\mathrm{MR} = 100\% \times [\rho(H) - \rho(0)]/\rho(0)$. For the temperatures well below $T_\mathrm{C}$, the negative MR is weak and a small enhancement in $\rho_{ab}(H)$ is observed when the field is low, which is also seen for FM-EuCd$_2$As$_2$ and should be related to the increased canting of spins towards the $c$ axis in the field~\cite{roychowdhuryAnomalousHallConductivity2023}. The normalized resistivity $\rho(H)/\rho(0\ \mathrm{T})$ at 72 K ($T_\mathrm{C}$), 85 K ($1.2T_\mathrm{C}$), and 130 K ($1.8T_\mathrm{C}$)
against normalized magnetization ($M/M_\mathrm{sat}$) is plotted in Fig. \ref{F3}(c). The resistivity at three different temperatures change with magnetization by following a similar trace. The good correlation implies that the MR of FM-EuZn$_2$P$_2$ is closely related to the magnetization, which means that magnetic scattering plays a major role in the MR of FM-EuZn$_2$P$_2$. In the inset of Fig. \ref{F3}(c), we plot $-$MR at 72 K, i.e., $-\Delta\rho/\rho(0\ \mathrm{T})$, as a function of $(M/M_\mathrm{sat})^2$. It is seen that the magnitude of $-$MR obeys the scaling function 
\begin{equation}
	-\Delta\rho/\rho(0\ \mathrm{T})=C_\mathrm{MR}(M/M_\mathrm{sat})^2,
\end{equation}
in the relatively low-$M$ region, say $M/M_\mathrm{sat}\le0.3$. The scaling factor $C_\mathrm{MR}$ is 2.85, resulting from the linear fit with the data in low-$M$ region. Since the metallic transport of FM-EuZn$_2$P$_2$, the Majumdar-Littlewood model is applicable to elucidate the dependence of MR on charge-carrier density~\cite{majumdarDependenceMagnetoresistivityChargecarrier1998}. According to the model, $C_\mathrm{MR}$ could be estimated with the relation $C_\mathrm{MR}\approx x^{-2/3}$, where $x$ is the number of charge carriers per magnetic unit cell. In the case of FM-EuZn$_2$P$_2$, $x\approx0.1$ due to 5\% Eu$^{2+}$ vacancies. Then we have $C_\mathrm{MR}=4.6$, comparable to the fitting value. Hence, the enhanced MR effect of FM-EuZn$_2$P$_2$ is expected by simply declining the carrier concentration, which could be realized through chemical doping or electrostatic gating.

We also measured the in-plane resistivity of AFM-EuZn$_2$P$_2$ for comparison, which is three orders of magnitude higher than that of FM-EuZn$_2$P$_2$ at 300 K, as shown in Fig. \ref{F3}(d). AFM-EuZn$_2$P$_2$ is insulating without the magnetic field. By applying a field of 9 T, $\rho_{ab}$ declines by half at 180 K yet the insulating behavior is retained. The band gap is estimated to be 0.42 eV with the Arrhenius model $\rho\propto e^{E_g/2k_\mathrm{B}T}$, consistent with earlier studies~\cite{berryTypeAntiferromagneticOrder2022,singhSuperexchangeInteractionInsulating2023a}. The distinct charge-transport behaviors of FM- and AFM-EuZn$_2$P$_2$ demonstrate that this system is successfully tuned from an AFM insulator into a FM metal by inducing the carriers.

\begin{figure}
	\includegraphics[width=0.47\textwidth]{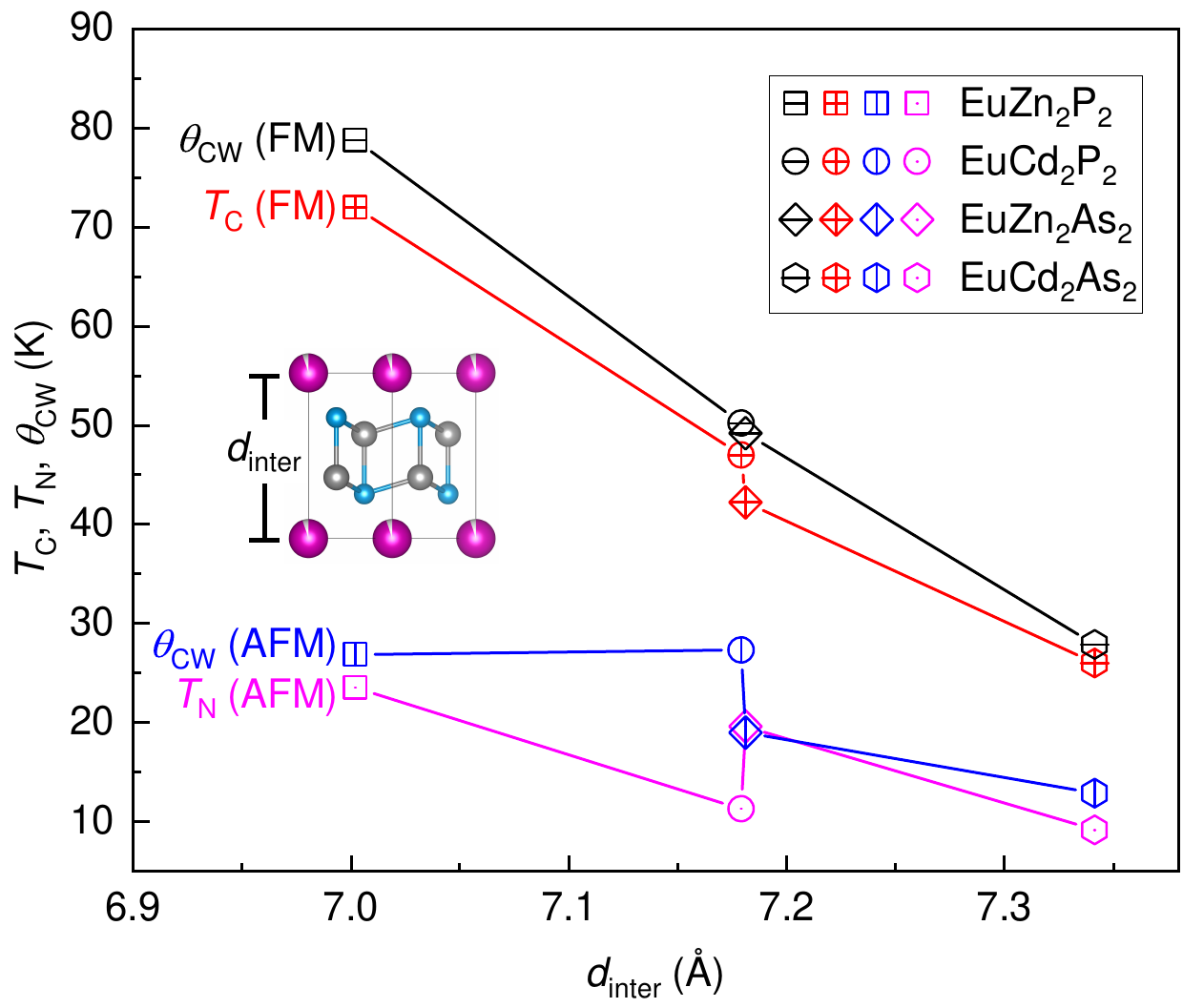}
	\caption{The characteristic temperatures ($T\rm_C$, $T\rm_N$, $\theta_\mathrm{CW}$) of Eu$M_2X_2$ ($M$ = Zn, Cd; $X$ = P, As) as a function of the Eu-layer distance, i.e., the $c$ axis. The squares, circles, diamonds, and hexagons represent the data of EuZn$_2$P$_2$, EuCd$_2$P$_2$, EuZn$_2$As$_2$, and EuCd$_2$As$_2$, respectively. The error bars are not shown because the uncertainties of data points are smaller than the size of symbols.}
	\label{F4}
\end{figure}

To get a deeper understanding of the carrier-induced ferromagnetism in Eu$M_2X_2$ ($M$ = Zn, Cd; $X$ = P, As), we synthesized FM-EuZn$_2$As$_2$ ($T_\mathrm{C}=42$ K) and FM-EuCd$_2$P$_2$ ($T_\mathrm{C}=47$ K) via the salt flux method as well. The characterizations of magnetism are presented in Figs. S3 and S4 in the SM~\cite{suppmatt}. The characteristic temperatures ($T_\mathrm{C}$, $T_\mathrm{N}$, $\theta_\mathrm{CW}$) of both FM- and AFM-Eu$M_2X_2$ are plotted as a function of the Eu-layer distance ($d_\mathrm{inter}$) in Fig. \ref{F4}. Note that since the structural difference between Eu$M_2X_2$ with the FM and AFM states is pretty small, we use the $c$-axis value reported for AFM-Eu$M_2X_2$ at room temperature as $d_\mathrm{inter}$ for both materials~\cite{berryTypeAntiferromagneticOrder2022,wangColossalMagnetoresistanceMixed2021,joManipulatingMagnetismTopological2020,wangAnisotropyMagneticTransport2022}. $T_\mathrm{N}$ and $\theta_\mathrm{CW}$ of AFM-Eu$M_2X_2$ do not show a monotonic dependence on $d_\mathrm{inter}$, while $T_\mathrm{C}$ and $\theta_\mathrm{CW}$ of FM-Eu$M_2X_2$ increase almost linearly with decreasing $d_\mathrm{inter}$. These linear relationships convincingly demonstrate that the FM ordering temperatures of FM-Eu$M_2X_2$ mainly depend on the interlayer Eu-Eu coupling. Thus $T_\mathrm{C}$ could be further enhanced by applying pressure or doping smaller divalent ions such as Ca$^{2+}$ to decrease the layer distance. Actually, $T_\mathrm{C}$ over 100 K was reported in pressurized EuZn$_2$As$_2$~\cite{luoColossalMagnetoresistanceTopological2023}. We notice that all values of $T_\mathrm{C}$ are slightly smaller than the corresponding $\theta_\mathrm{CW}$, which is ascribed to the influence of existing interlayer AFM interaction through the Eu-$X$-$X$-Eu superexchange path, as mentioned earlier~\cite{singhSuperexchangeInteractionInsulating2023a}. Therefore, it is natural to see a larger discrepancy between $T_\mathrm{C}$ and $\theta_\mathrm{CW}$ for FM-EuZn$_2$P$_2$ and FM-EuZn$_2$As$_2$ due to their shorter Eu-$X$-$X$-Eu path.

Our results manifest that the interlayer FM coupling is essential to the FM ordering temperature, and the FM interaction results from the indirect exchange mediated by carriers. The role of carrier densities has not been discussed yet. It seems that the influence of carrier concentration on $T_\mathrm{C}$ is not primary in the case of Eu$M_2X_2$. For example, $T_\mathrm{C}$ and $\theta_\mathrm{CW}$ of FM-EuCd$_2$As$_2$ vary only several kelvins when the chemical doping level changes by an order of magnitude~\cite{kuthanazhiMagnetismPhaseDiagrams2023}. Since the carrier concentrations of FM-Eu$M_2X_2$ are lower than the validity condition of the Ruderman-Kittel-Kasuya-Yosida (RKKY) theory ($>10^{21}\ \mathrm{cm}^{-3}$), the dependence of interlayer FM interaction on the carrier density cannot be understood within this framework~\cite{Nagaev1986}. Nevertheless, the correlation between $T_\mathrm{C}$ and carrier density is not ruled out. Further theoretic and experimental efforts will be devoted to investigating the effect of carrier density on $T_\mathrm{C}$.

In summary, FM-EuZn$_2$P$_2$ was successfully grown via the salt flux method, which has similar structural parameters to AFM-EuZn$_2$P$_2$ except for 5\% Eu vacancies. The magnetization and resistivity measurements show that the defective EuZn$_2$P$_2$ is a FM metal of $T_\mathrm{C}$ = 72 K, rather than an AFM insulator of $T_\mathrm{N}$ = 23.5 K, as in the vacancy-free version. The analysis of field-dependent resistivity indicates that the magnetic scattering makes the main contribution to the resistivity near $T_\mathrm{C}$. On basis of the Majumdar-Littlewood model, a greater MR effect is probably achieved by reducing the carrier density. The transition temperatures and Curie-Weiss temperatures of AFM- and FM-Eu$M_2X_2$ ($M$ = Zn, Cd; $X$ = P, As) are examined, which reveals linear dependences of $\theta_\mathrm{CW}$ and $T_\mathrm{C}$ on the Eu-layer distance, indicating the critical role of interlayer FM coupling on the FM ordering. Our results suggest that FM coupling could be induced with low carrier densities in Eu-based CaAl$_2$Si$_2$-type materials, which may also be applicable to other AFM Eu-based Zintl compounds such as EuIn$_2$As$_2$ and EuMn$_2$P$_2$, for their narrow energy gaps and similar layered structures. Moreover, it is of great interest to explore other exotic phenomena such as the topological phase transition by tuning the carrier concentration of Eu$M_2X_2$ in controlled manners like chemical doping and the gating technique, which is not only important for fundamental research, but also significant for potential applications in spintronics.

\begin{acknowledgments}
This work was supported by the National Natural Science Foundation of China (Grants No. 12204094), the Natural Science Foundation of Jiangsu Province (Grant No. BK20220796), the Start-up Research Fund of Southeast University (Grant No. RF1028623289), the Interdisciplinary program of Wuhan National High Magnetic Field Center (WHMFC) at Huazhong University of Science and Technology (Grant No. WHMFC202205), and the open research fund of Key Laboratory of Quantum Materials and Devices (Southeast University), Ministry of Education.
\end{acknowledgments}

\bibliography{FMEZP}
\end{document}

%\preprint{APS/123-QED}

\title{Supplemental Materials: Carrier-induced transition from antiferromagnetic insulator to ferromagnetic metal in the layered phosphide EuZn$_2$P$_2$}

\author{Xiyu Chen}
\affiliation{Key Laboratory of Quantum Materials and Devices of Ministry of Education, School of Physics, Southeast University, Nanjing 211189, China}
\author{Wuzhang Yang}
\affiliation{School of Science, Westlake University, Hangzhou 310024, China}
\affiliation{Institute of Natural Sciences, Westlake Institute for Advanced Study, Hangzhou 310024, China}
\author{Jia-Yi Lu}
\affiliation{School of Physics, Interdisciplinary Center for Quantum Information and State Key Laboratory of Silicon and Advanced Semiconductor Materials, Zhejiang University, Hangzhou 310058, China}
\author{Zhiyu Zhou}
\affiliation{Key Laboratory of Quantum Materials and Devices of Ministry of Education, School of Physics, Southeast University, Nanjing 211189, China}
\author{Zhi Ren}
\affiliation{School of Science, Westlake University, Hangzhou 310024, China}
\affiliation{Institute of Natural Sciences, Westlake Institute for Advanced Study, Hangzhou 310024, China}
\author{Guang-Han Cao}
\affiliation{School of Physics, Interdisciplinary Center for Quantum Information and State Key Laboratory of Silicon and Advanced Semiconductor Materials, Zhejiang University, Hangzhou 310058, China}
\affiliation{Collaborative Innovation Centre of Advanced Microstructures, Nanjing University, Nanjing 210093, China}
\author{Shuai Dong}
%\email{sdong@seu.edu.cn}
\affiliation{Key Laboratory of Quantum Materials and Devices of Ministry of Education, School of Physics, Southeast University, Nanjing 211189, China}
\author{Zhi-Cheng Wang}
\email{wzc@seu.edu.cn}
\affiliation{Key Laboratory of Quantum Materials and Devices of Ministry of Education, School of Physics, Southeast University, Nanjing 211189, China}


\date{\today}% It is always \today, today,

\maketitle

\begin{center}
	\textbf{Content}
	\begin{description}
		\item[A] Experimental methods for the crystal growth and characterizations.
		\item[B] Figure S1 and Table S1. Chemical composition analysis of FM-EuZn$_2$P$_2$ via energy dispersive spectroscopy.
		\item[C] Figure S2 and Table S2. Chemical composition analysis of AFM-EuZn$_2$P$_2$ via energy dispersive spectroscopy.
		\item[D] Figure S3 and Table S3. X-ray diffraction pattern and the Rietveld refinement of AFM-EuZn$_2$P$_2$
		
%		\item Figure S2: Comparison of $T_\mathrm{C}$ between EuZn$_2$P$_2$ samples from different batches.
		\item[E] Figure S4 and S5. Magnetic data of FM- and AFM-EuZn$_2$P$_2$
		\item[F] Figure S6. Magnetism of FM-EuZn$_2$As$_2$
		\item[G] Figure S7. Magnetism of FM-EuCd$_2$P$_2$
		\item[H] Figure S8. SEM images and chemical composition of FM-EuZn$_2$As$_2$
	\end{description}
\end{center}
\clearpage

\subsection{Methods}

\textit{Crystal growth.} The Eu ingots (99.999\%), Zn powder (99.99\%), red P lumps (99.999\%), As powder (99.99\%), Sn shots (99.99\%), NaCl (99.99\%), and KCl (99.99\%) were used to grow single crystals of EuZn$_2$P$_2$ and EuZn$_2$As$_2$. FM-EuZn$_2$P$_2$ single crystals were grown with salt flux of an equimolar mixture of NaCl/KCl. A total mass of 1 g of Eu, Zn, and P was weighed using a molar ratio of 1:2:2. The reactants were mixed with 4 g of the salt flux, then placed in a quartz tube, which was sealed under high vacuum ($<10^{-2}$ Pa). Next, the firstly sealed quartz tube was loaded into a second quartz tube with slightly larger diameter, which was also vacuumed and sealed. The nested tubes were heated to 469 $^\circ$C over 24 h, held for 24 h, then heated to 597 $^\circ$C in a rate of 20 $^\circ$C/h and held for another 24 h. Afterwards, the tubes were heated to 847 $^\circ$C and held for 50 h, and subsequently cooled to 630 $^\circ$C at 1.5 $^\circ$C/h. Shiny flakes of crystals can be obtained by washing the product with deionized water. FM-EuZn$_2$As$_2$ and FM-EuCd$_2$P$_2$ single crystals were grown through a similar method like FM-EuZn$_2$P$_2$.
%However, the size and quality of FM-EuZn$_2$As$_2$ single crystals were not good enough for the measurements of transport properties. 
The scanning electron microscopy (SEM) micrographs of FM-EuZn$_2$As$_2$ are shown in Fig. \ref{fig:FigS5}, which are brittle and not good enough for the measurements of transport properties. AFM-EuZn$_2$P$_2$ was grown by Sn flux. The elements were mixed in a molar ratio Eu:Zn:P:Sn = 1:1:1:8. The mixture was loaded into an alumina crucible inside an evacuated quartz ampule and slowly heated to 1100 $^\circ$C, held for 24 h, cooled to 900 $^\circ$C at 3 $^\circ$C/h, cooled to 600 $^\circ$C at 5 $^\circ$C/h, and finally centrifuged to remove the flux.

%It is worth noting that we also tried the growth of EuZn$_2$P$_2$ in a ratio of Eu:Zn:P = 0.9:2:2. The transition temperature of harvested crystals are close to the samples in the main text, which is also presented in the Supplemental Materials (Fig. \ref{fig:FigS2}).

\textit{Structure Determination and Chemical Compositions Characterization.} The PANalytical x-ray diffractometer with the Cu $K\alpha$1 radiation was used to check the quality of single crystals at room temperature. The crystal structure of FM-EuZn$_2$P$_2$ at 150 K was determined by a Bruker D8 Venture diffractometer equipped with I$\mu$S 3.0 Dual Wavelength system (Mo $K$$\alpha$ radiation, $\lambda$ = 0.71073 Å), and an APEX-II CCD detector. The initial structural model was developed with the intrinsic phasing feature of SHELXT and a least-square refinement was performed using SHELXL2014. The chemical compositions of single crystals were characterized by scanning electron microscopy (FEI Inspect F50) with Amatek EDAX EDS detector.

\textit{Magnetization and Transport Measurements.} The direct-current (dc) magnetization was measured using a magnetic property measurement system (MPMS-3, Quantum Design). The resistivity was collected with a standard four-probe technique using a physical property measurement system (PPMS-9, Quantum Design).

%\clearpage

\subsection{Chemical composition analysis of FM-EuZn$_2$P$_2$}

%%%%%%%%%%%%%%%%%% FIGURES3 %%%%%%%%%%%%%%%%%%%%%
\begin{figure}[htbp]
	\centering
	\includegraphics[width=0.75\columnwidth]{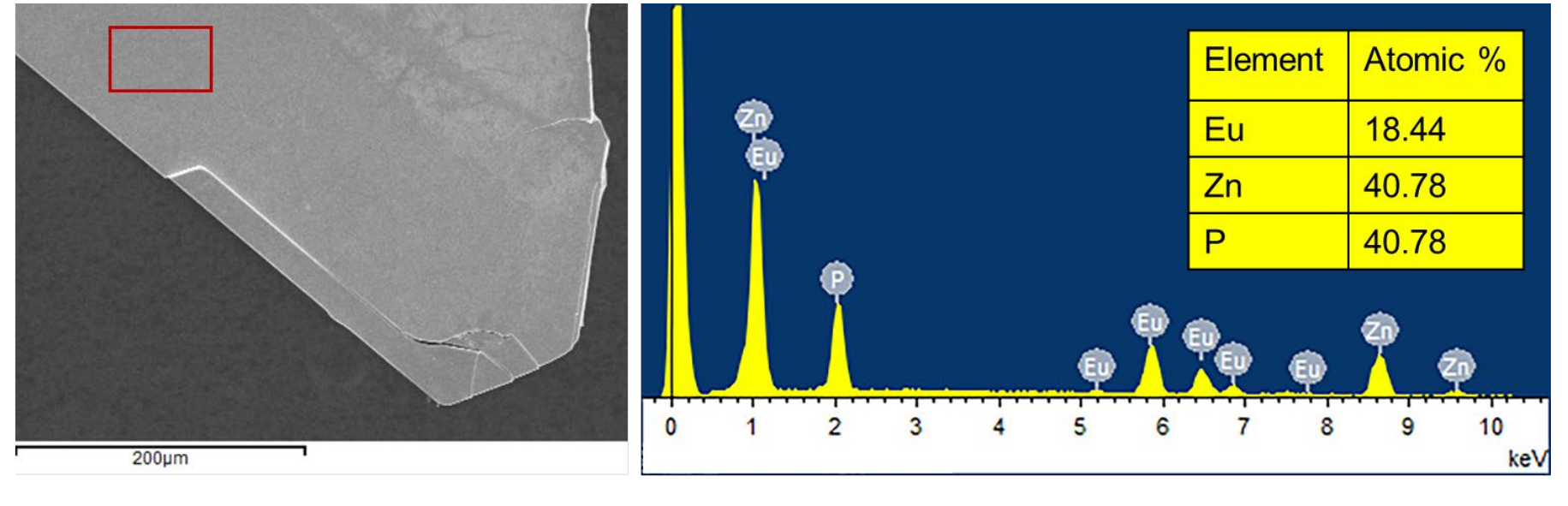}
	\caption{A typical energy dispersive x-ray spectrum of FM-EuZn$_2$P$_2$ single crystal.}
	\label{fig:FigEDX}
\end{figure}
%%%%%%%%%%%%%%%%%% FIGURES3 %%%%%%%%%%%%%%%%%%%%%

	%%%%%%%%%%%%%%%%%%% TABLES1 %%%%%%%%%%%%%%%%%%%%%
\begin{table}[htbp]
	\caption{\label{TabS1} Chemical composition analysis of FM-EuZn$_2$P$_2$ single crystals via energy dispersive x-ray spectroscopy. The number of Zn atoms in a formula unit is normalized to 2, and other data are rescaled accordingly. The first column shows the data points and corresponding specimens, for example, "S1, P1" represents the first point from sample 1, and so on. The resulting chemical formula is Eu$_{0.96(3)}$Zn$_{2.00}$P$_{2.11(6)}$.}
%	\resizebox{0.5\textwidth}{!}{
%	\begin{ruledtabular}

		\begin{tabular}{p{3cm}<{\centering}|p{2cm}<{\centering}|p{2cm}<{\centering}|p{2cm}<{\centering}|p{2cm}<{\centering}}
			\hline\hline
				& Eu & Zn &  P  & Eu/P \\
				\hline
			S1, P1&	0.93 &	2.00 &	2.15  & 0.43\\
			S1, P2&	1.00 &	2.00 &	2.22  & 0.45 \\
			\hline
			S2, P1&	0.93 &	2.00 &	2.11  & 0.44\\
			S2, P2&	0.94 &	2.00 &	2.05  & 0.46\\
			S2, P3&	0.90 &	2.00 &	2.00  & 0.45\\
			\hline
			S3, P1&	0.96 &	2.00 &	2.07  & 0.46\\
			S3, P2&	0.97 &	2.00 &	2.13  & 0.45\\
			\hline
			S4, P1&	0.99 &	2.00 &	2.13  & 0.46\\
			S4, P2&	0.99 &	2.00 &	2.08  & 0.48\\
			S4, P3&	1.00 &	2.00 &	2.11  & 0.47\\
			\hline
			average	& 0.96 &	2.00 &	2.11  & 0.46\\
			
			stand deviation&0.03 &	0.00 &	0.06  & 0.013			\\
			\hline\hline
		\end{tabular}
	
%	\end{ruledtabular}

\end{table}
%c{2cm}<{\centering}c{2cm}<{\centering}c{2cm}<{\centering}

\clearpage

\subsection{Chemical composition analysis of AFM-EuZn$_2$P$_2$}

%%%%%%%%%%%%%%%%%% FIGURES3 %%%%%%%%%%%%%%%%%%%%%
\begin{figure}[htbp]
	\centering
	\includegraphics[width=0.75\columnwidth]{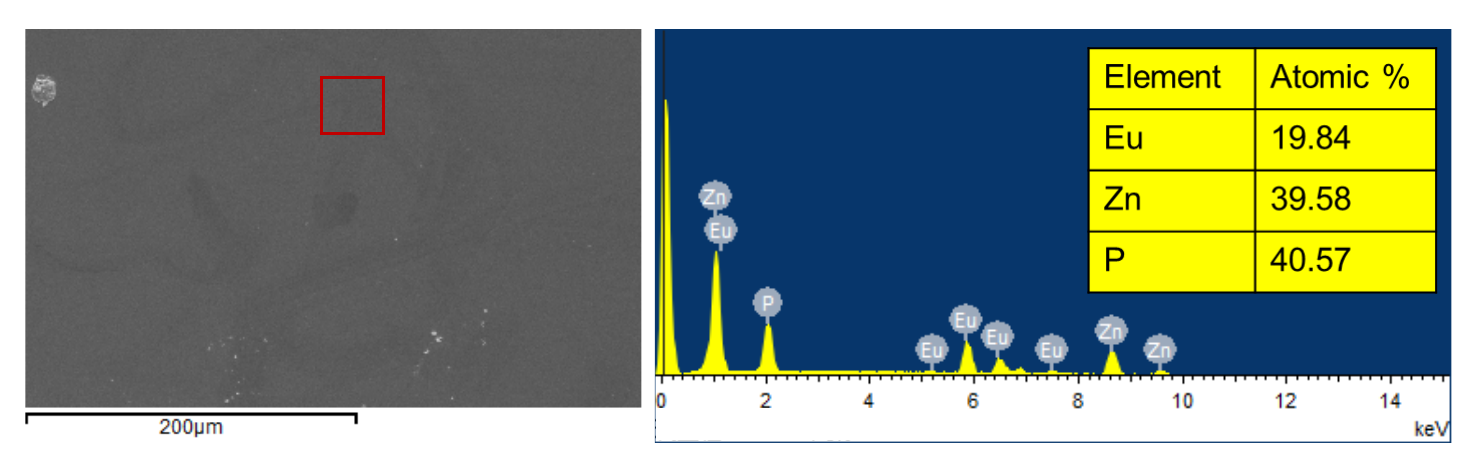}
	\caption{A typical energy dispersive x-ray spectrum of AFM-EuZn$_2$P$_2$ single crystal.}
	\label{fig:FigEDX2}
\end{figure}
%%%%%%%%%%%%%%%%%% FIGURES3 %%%%%%%%%%%%%%%%%%%%%

%%%%%%%%%%%%%%%%%%% TABLES1 %%%%%%%%%%%%%%%%%%%%%
\begin{table}[htbp]
	\caption{\label{TabS2} Chemical composition analysis of AFM-EuZn$_2$P$_2$ single crystals via energy dispersive x-ray spectroscopy. The number of Zn atoms in a formula unit is normalized to 2, and other data are rescaled accordingly. The resulting chemical formula is Eu$_{0.99(2)}$Zn$_{2.00}$P$_{1.98(13)}$.}.
	%	\resizebox{0.5\textwidth}{!}{
		%	\begin{ruledtabular}
			
			\begin{tabular}{p{3cm}<{\centering}|p{2cm}<{\centering}|p{2cm}<{\centering}|p{2cm}<{\centering}|p{2cm}<{\centering}}
				\hline\hline
				& Eu & Zn &  P & Eu/P\\
				\hline

				S1, P1&1.00&	2.00	&2.05 & 0.49\\
				S1, P2&1.02&	2.00&	2.13 & 0.48\\
				S1, P3&1.00&	2.00&	2.13 & 0.47\\
				S1, P4&1.01	&2.00&	2.11 & 0.48 \\
				\hline
				S2, P1&0.96&	2.00&	1.89 & 0.51\\
				S2, P2&0.97&	2.00&	1.85 & 0.52\\
				S2, P3&0.98&	2.00&	1.84 & 0.53\\
				S2, P4&0.99&	2.00&	1.88 & 0.53\\

				\hline
				average	& 0.99 &	2.00 &	1.98 &  0.50 \\
				
				stand deviation&0.02 &	0.00 &	0.13 & 0.03 \\
				\hline\hline
			\end{tabular}
			
			%	\end{ruledtabular}
		
	\end{table}

\clearpage
\subsection{Structural analysis of AFM-EuZn$_2$P$_2$}

%%%%%%%%%%%%%%%%%% FIGURES3 %%%%%%%%%%%%%%%%%%%%%
\begin{figure}[htbp]
	\centering
	\includegraphics[width=0.8\columnwidth]{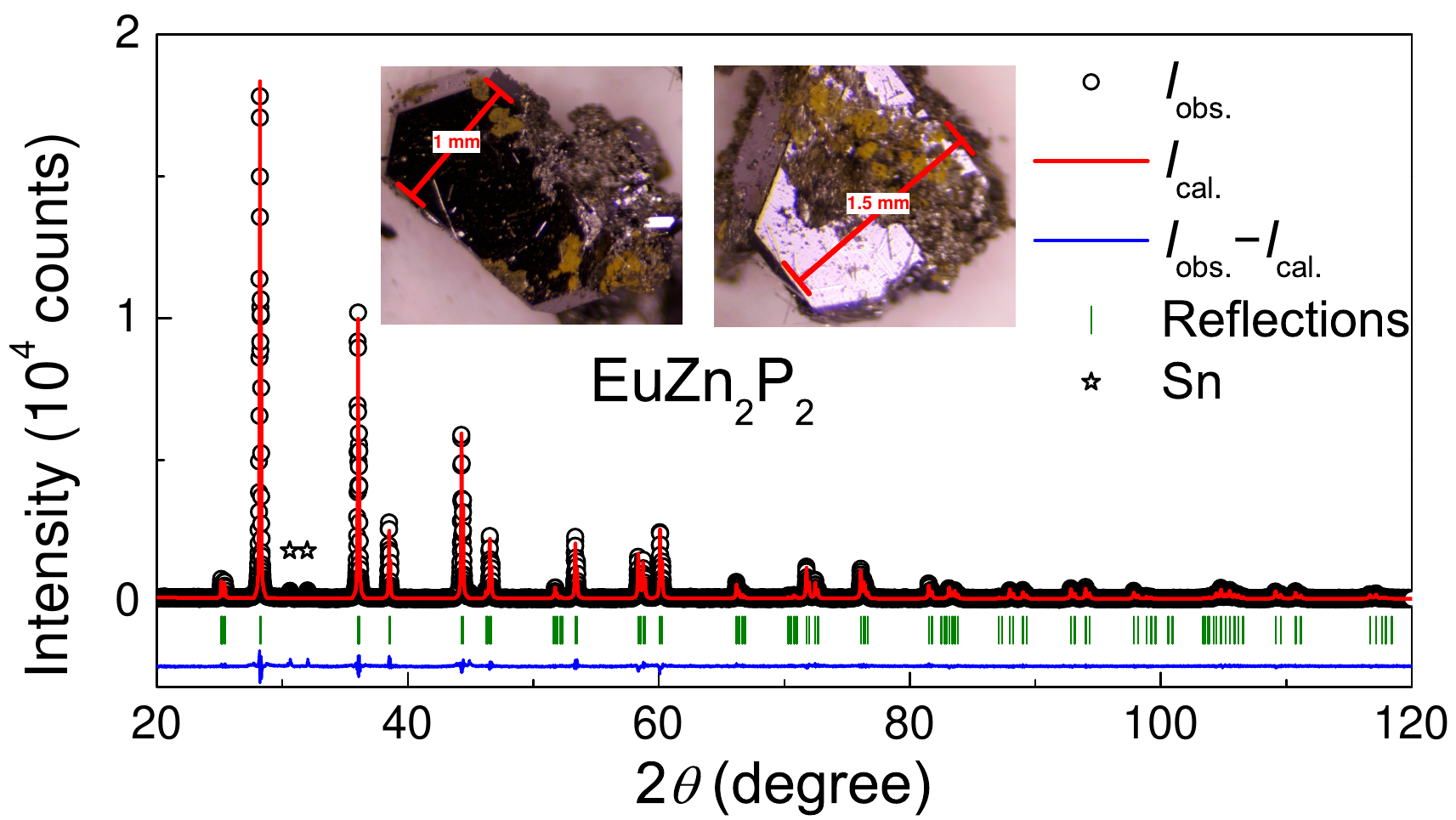}
	\caption{X-ray diffraction pattern and the Rietveld refinement of AFM-EuZn$_2$P$_2$, collected at room temperature by grinding the single crystals into powder.}
	\label{fig:FigXRD}
\end{figure}
%%%%%%%%%%%%%%%%%% FIGURES3 %%%%%%%%%%%%%%%%%%%%%

%%%%%%%%%%%%%%%%%%% TABLES1 %%%%%%%%%%%%%%%%%%%%%
\begin{table}[htbp]
	\caption{\label{TabS3} Unit cell dimensions and refinement parameters are listed for AFM-EuZn$_2$P$_2$ from the Rietveld refinements in the space group $P\bar{3}m1$. The atomic coordinates are as follows: Eu $1a$ (0,0,0); Zn $2d$ (1/3,2/3,0.6310(2)); P $2d$ (1/3,2/3,0.2706(4)). All sites are fully occupied.}
	%	\resizebox{0.5\textwidth}{!}{
		%	\begin{ruledtabular}
			
			\begin{tabular}{ll}
				\hline\hline
				Material & ~~~~~~~~AFM-EuZn$_2$P$_2$ \\
				$a$ (\AA) &  ~~~~~~~~4.08685(2)\\
				$c$ (\AA) &  ~~~~~~~~7.00784(5)\\
				$c/a$ &  ~~~~~~~~1.7147\\
				$V$ (\AA$^3$) &  ~~~~~~~~101.366(2)\\
				Goodness-of-fit &  ~~~~~~~~1.44 \\
				$R_\mathrm{wp}$ &  ~~~~~~~~8.41 \\
				$R_\mathrm{exp}$ &  ~~~~~~~~5.83 \\
				$R_\mathrm{p}$ &  ~~~~~~~~6.52 \\
				\hline\hline
			\end{tabular}
			
			%	\end{ruledtabular}
		
	\end{table}
\clearpage

%\clearpage
\subsection{Magnetic data of FM- and AFM-EuZn$_2$P$_2$}
%\subsection{$M(H)$ curves and Curie-Weiss fit of $\chi(T)$ with out-of-plane fields for FM-EuZn$_2$P$_2$ }

%%%%%%%%%%%%%%%%%% FIGURES3 %%%%%%%%%%%%%%%%%%%%%
\begin{figure}[htbp]
	\centering
	\includegraphics[width=0.9\columnwidth]{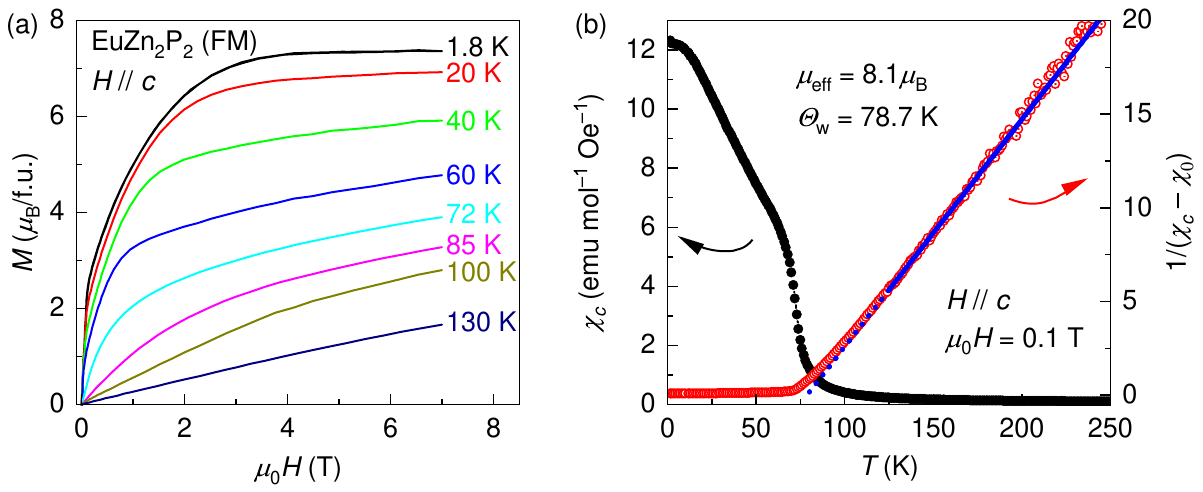}
	\caption{(a) $M(H)$ curves of FM-EuZn$_2$P$_2$ with $H\parallel c$. (b) Curie-Weiss analysis from 125 to 300 K with $\chi_{c}(T)$ data at 0.1 T.}
	\label{fig:FigS1}
\end{figure}
%%%%%%%%%%%%%%%%%% FIGURES3 %%%%%%%%%%%%%%%%%%%%%

%\clearpage
%\subsection{Magnetic data of AFM-EuZn$_2$P$_2$}

%%%%%%%%%%%%%%%%%% FIGURES4 %%%%%%%%%%%%%%%%%%%%%
\begin{figure}[htbp]
	\centering
	\includegraphics[width=0.9\columnwidth]{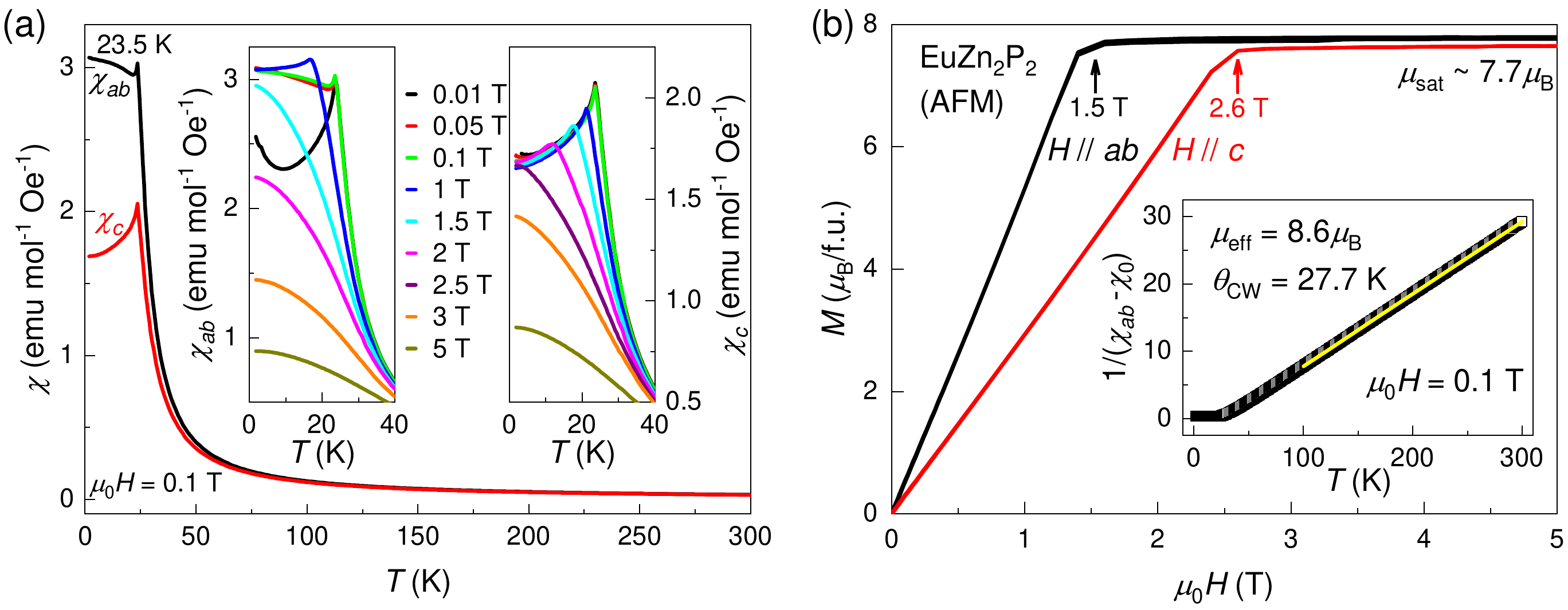}
	\caption{(a) Temperature dependences of anisotropic magnetic susceptibility for AFM-EuZn$_2$P$_2$. $\chi_{ab}(T)$ and $\chi_{c}(T)$ at 0.1 T are compared in the main panel, while the insets show $\chi(T)$ curves with different fields. (b) Anisotropic magnetizations as a function of field at 1.8 K. Inset shows the Curie-Weiss fit with the in-plane susceptibility.}
	\label{fig:FigS2}
\end{figure}
%%%%%%%%%%%%%%%%%% FIGURES4 %%%%%%%%%%%%%%%%%%%%%
\clearpage
\subsection{Magnetism of FM-EuZn$_2$As$_2$}

%%%%%%%%%%%%%%%%%% FIGURES5 %%%%%%%%%%%%%%%%%%%%%
\begin{figure}[htbp]
	\centering
	\includegraphics[width=0.9\columnwidth]{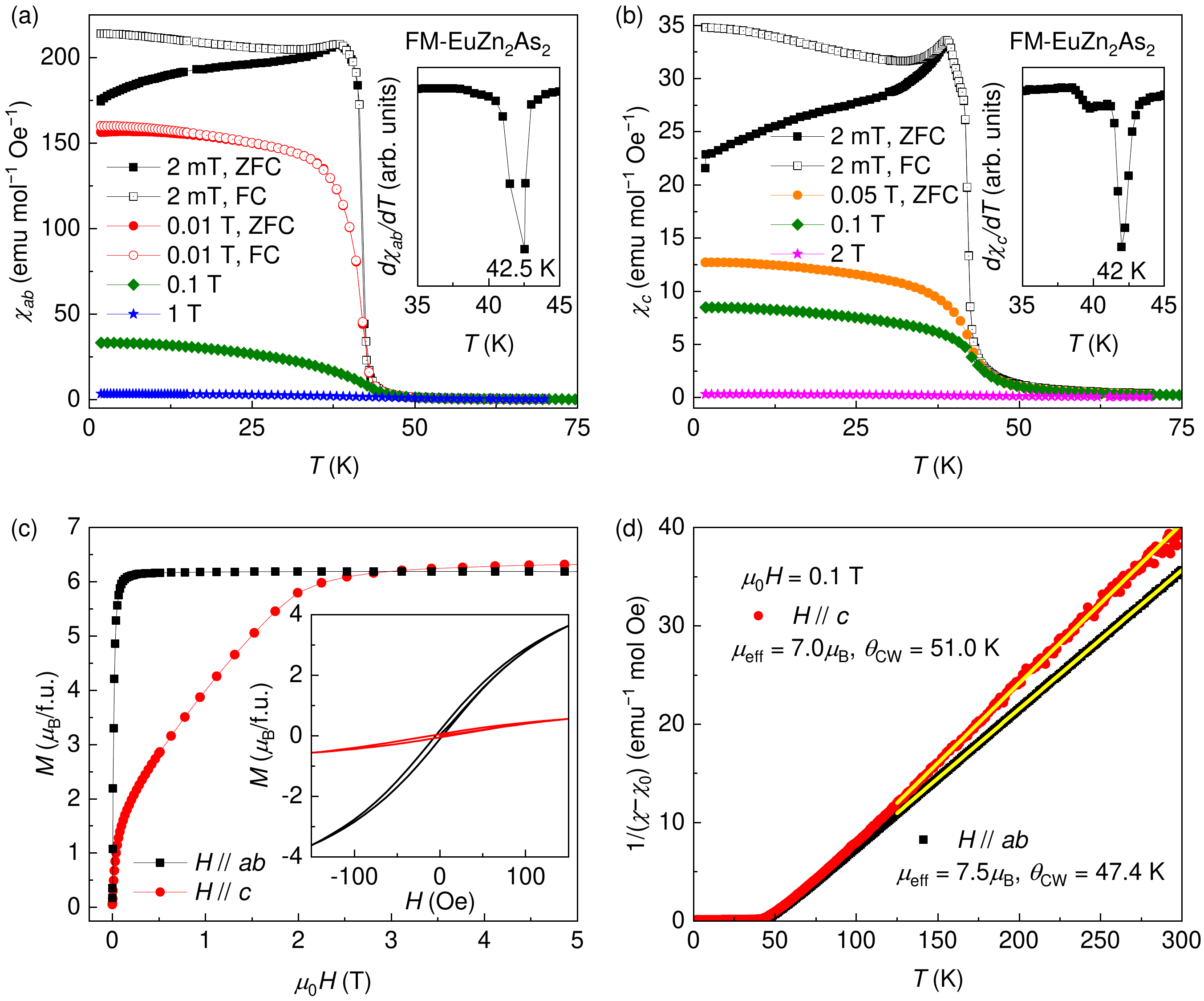}
	\caption{[(a),(b)] Temperature dependence of magnetic susceptibility with the field in the $ab$ plane and along the $c$ axis. Insets present the transition points determined by the derivative ($d\chi/d T$). (c) Anisotropic magnetizations as a function of field for FM-EuZn$_2$As$_2$. Inset shows the hysteresis loops under the low external fields. (d) Curie-Weiss analysis of $\chi(T)$.}
	\label{fig:FigS3}
\end{figure}
%%%%%%%%%%%%%%%%%% FIGURES5 %%%%%%%%%%%%%%%%%%%%%
\clearpage
\subsection{Magnetism of FM-EuCd$_2$P$_2$}

%%%%%%%%%%%%%%%%%% FIGURES6 %%%%%%%%%%%%%%%%%%%%%
\begin{figure}[htbp]
	\centering
	\includegraphics[width=0.9\columnwidth]{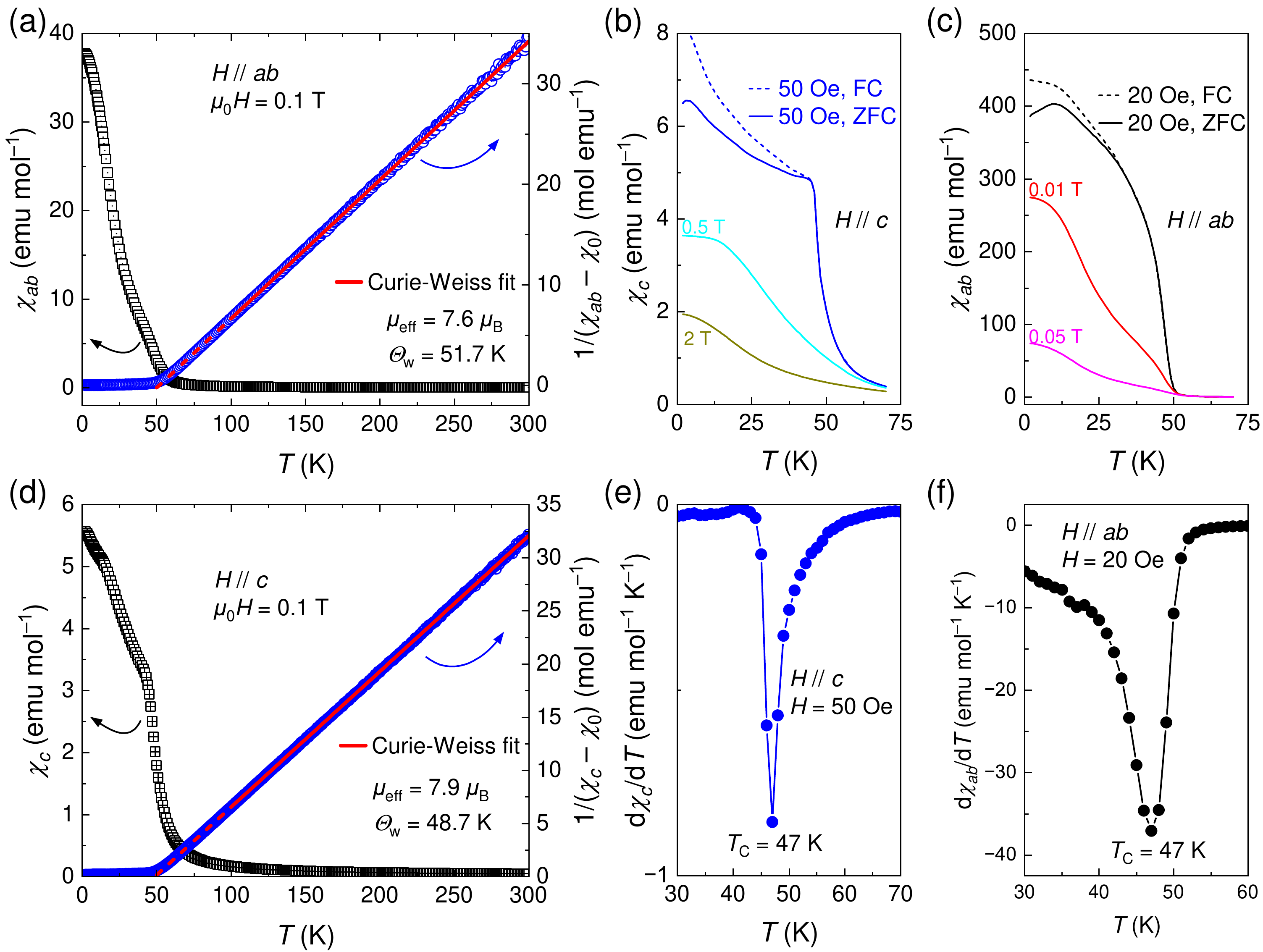}
	\caption{[(a),(d)] The in-plane and out-of-plane susceptibility measured under 0.1 T (left axes), and the corresponding reciprocal of susceptibility as well as the Curie-Weiss fits from 100 to 300 K (right axes). [(b),(c)] Temperature dependences of magnetic susceptibility $\chi(T)$ in various out-of-plane and in-plane fields. [(e),(f)] Transition temperatures determined by $d\chi_{c}/d T$ and $d\chi_{ab}/d T$, respectively.}
	\label{fig:FigS4}
\end{figure}
\clearpage

\subsection{SEM images and compositions of FM-EuZn$_2$As$_2$}

%%%%%%%%%%%%%%%%%% FIGURES1 %%%%%%%%%%%%%%%%%%%%%
\begin{figure}[htbp]
	\centering
	\includegraphics[width=0.8\columnwidth]{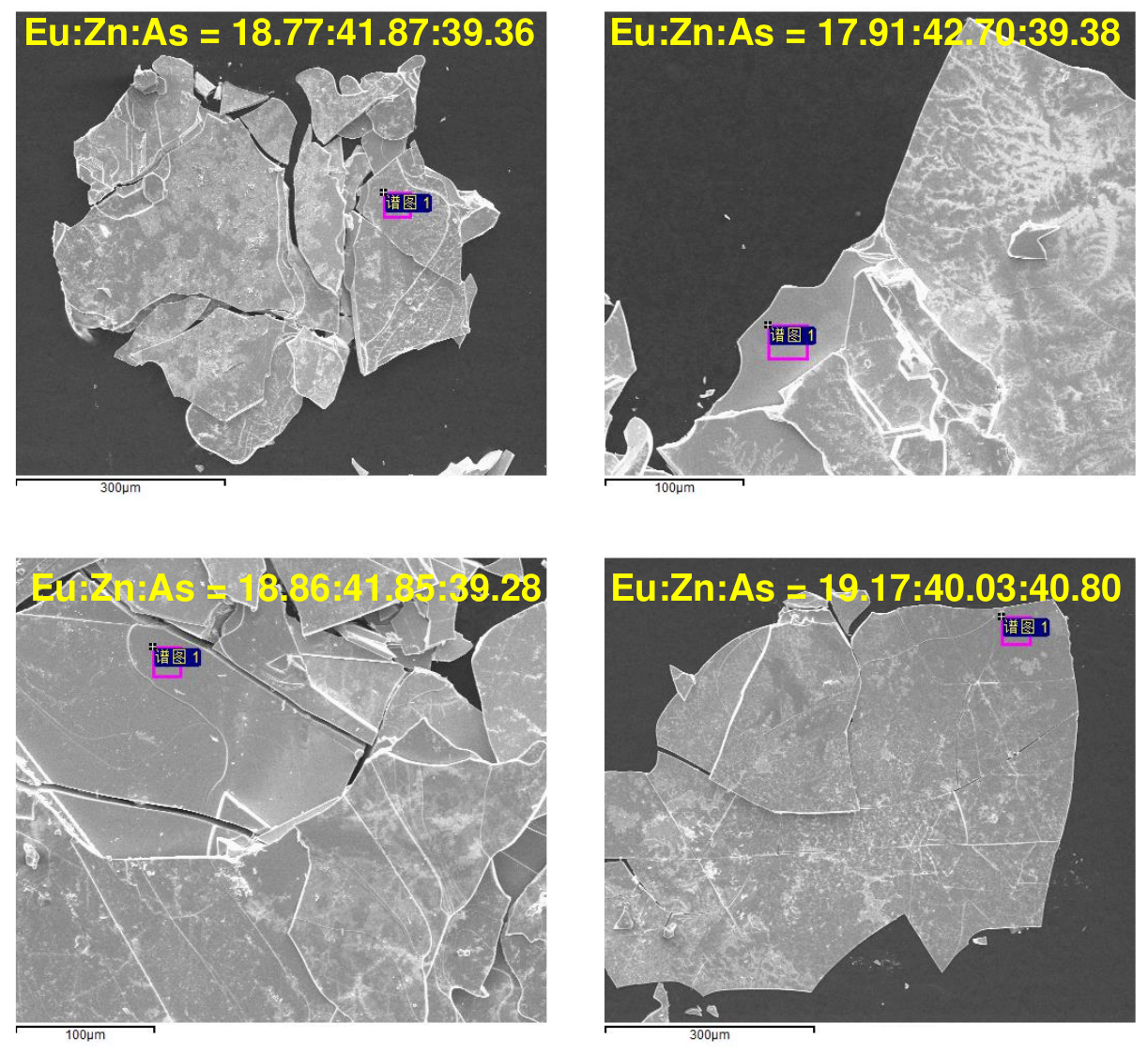}
	\caption{The SEM images of FM-EuZn$_2$As$_2$ single crystals. The compositions collected by EDX are shown as yellow text on the images. The crystals are brittle.}
	\label{fig:FigS5}
\end{figure}
%%%%%%%%%%%%%%%%%% FIGURES1 %%%%%%%%%%%%%%%%%%%%%
\clearpage

%%%%%%%%%%%%%%%%%% FIGURES5 %%%%%%%%%%%%%%%%%%%%%
%\subsection{Comparison of $T_\mathrm{C}$ between EuZn$_2$P$_2$ samples from different batches}
%
%
%%%%%%%%%%%%%%%%%%% FIGURES2 %%%%%%%%%%%%%%%%%%%%%
%\begin{figure}[htbp]
%	\centering
%	\includegraphics[width=0.5\columnwidth]{FigS2}
%	\caption{FM transitions of EuZn$_2$P$_2$ samples from different batches. Batch A is synthesized with the element ratio of Eu:Zn:P = 1:2:2, while Batch B is synthesized with Eu:Zn:P = 0.9:2:2. $T_\mathrm{C}$ of Batch B is several kelvins lower. The magnetic and resistivity data in the main text are all collected with crystals from Batch A.}
%	\label{fig:FigS2}
%\end{figure}
%%%%%%%%%%%%%%%%%%% FIGURES2 %%%%%%%%%%%%%%%%%%%%%

%\bibliography{}% Produces the bibliography via BibTeX.